\documentclass{emulateapj}

\usepackage{amsmath}
\usepackage{graphicx} 

\begin{document}
\shorttitle{Ultra-steep spectrum giant radio galaxy 
in Abell\,449
}
\shortauthors{Hunik \& Jamrozy}

\title{%
Discovery of ultra-steep spectrum giant radio galaxy with recurrent radio jet activity in Abell\,449
}

\author{
Dominika Hunik $^{1,2}$
and 
Marek Jamrozy $^{1}$
}

\affil{
$^{1}$ Obserwatorium Astronomiczne, Uniwersytet Jagiello\'nski, ul. Orla 171, 30-244 Krak\'ow, Poland; dominika.hunik@uj.edu.pl
\\
$^{2}$ Instytut Fizyki, Uniwersytet Jagiello\'nski, ul. {\L}ojasiewicza 11, 30-348 Krak\'ow, Poland
}

\begin{abstract}
We report a discovery of a 1.3\,Mpc diffuse radio source with extremely steep spectrum fading radio 
structures in the vicinity of the Abell\,449 cluster of galaxies. Its extended diffuse lobes are 
bright only at low radio frequencies and their synchrotron age is about 160\,Myr. The parent galaxy 
of the extended relic structure, which is the dominant galaxy within the cluster, is starting a new 
jet activity. There are three weak X-rays sources in the vicinity of the cluster as found in the ROSAT 
survey, however it is not known if they are connected with this cluster of galaxies. 
Just a few radio galaxy relics are currently known in the literature, as finding them requires sensitive and high angular resolution low-frequency radio observations. Objects of this kind, which also are starting a new jet activity, are important for understanding the life cycle and evolution of active galactic 
nuclei. A new 613\,MHz map as well as the archival radio data pertaining to this object are presented 
and analyzed.
\end{abstract}

\section{Introduction}
\label{sec:intro}

Clusters of galaxies are ideal laboratories to probe the evolution and formation of cosmic 
structures. Apart from luminous matter, they consist of intracluster medium (ICM), 
which can be studied through X-ray astronomy (e.g., \citealt{Sarazin1986}). 
The ICM is mixed with non-thermal components, like relativistic 
particles or large-scale magnetic fields, which emit synchrotron radiation detected at radio wavelengths. 
These sources are typically grouped into three kinds of objects: halos, mini-halos, 
relics~(\citealt{Ferrari2008, Venturi2011, Feretti2012}, and  references  therein). The halos 
usually extend to over 1\,Mpc, have a regular morphology, low surface brightness and are unpolarized. 
They always show a steep spectrum, i.e. their spectral index $\alpha\gtrsim 1$ ($S_{\nu}\propto\nu^{-\alpha}$). The mini-halos are structures extending to about 500\,kpc, also 
with a steep spectrum and low surface brightness. The relics are similar to halos in their surface brightness, 
large size, and spectral index, but they are located in clusters' peripheral regions and strongly polarized. 
Most of them are elongated, but roundish relics can also be observed.

In the ICM, there are also jets and lobes of radio galaxies. During their 
active stage, the powerful extragalactic radio sources are supplied with energy from the active galactic 
nucleus (AGN) in the form of jets or plasma beams. The structure and spectral index distribution of the lobes 
of extended radio emission contain important information on the history of the source. For young sources, the 
radio spectra are usually well approximated as a power law with a spectral index about 0.5. When the jet activity ceases, energy is no longer delivered, and the 
structures gradually fade away due to adiabatic and radiative losses of relativistic electrons, before completely disappearing. Such relic radio galaxies are essential for understanding the radio source evolution, in particular, 
the late phase of exhaustion of the central energy source, AGN. Dying diffuse sources, arising from the AGNs after stopping the nuclear activity are sometimes referred as relics or AGN relics, even though 
they have not originated from the ICM. The spectrum of the aged sources can be described by a power law with a high-energy cut-off. In effect, it becomes steep with an inflection at a break frequency. The break frequency is 
associated with the synchrotron age of charged particles. The first example of a dying radio source was 
B2\,0924$+$30~\citep{Cordey1987, Jamrozy2004}. Nine dying sources, including three with recurrent activity, 
were discovered using the Westerbork Northern Sky Survey (WENSS; 326\,MHz; ~\citealt{Rengelink1997}) and the 
NRAO VLA Sky Survey (NVSS;~\citealt{Condon1998}) by~\cite{Parma2007}. \cite{Murgia2011} described other five 
sources of this class. The radio structures within those objects vary in size from about 7 to 135\,kpc, while 
recently, \cite{Brienza2016} described a new remnant radio galaxy, with a physical extent of 700\,kpc.

In this paper, we present new observations carried out with the Giant Metrewave Radio Telescope (GMRT) at 
613\,MHz as well as archival data on the ultra-steep spectrum giant radio galaxy J0349$+$7511 with hints of 
recurrent jet activity. In the next section, we briefly describe the radio source and its environment. 
The third section presents the GMRT observations as well as available archival data from other radio sky surveys. 
The archival infrared and X-ray data are described in the fourth section, while in the fifth section we review 
and discuss the results. The conclusions are given at the end of the paper. Throughout the letter, a flat 
vacuum-dominated universe with $\Omega$$\rm_{m}$=0.27, $\Omega$$\rm_{\Lambda}$=0.73 and 
$H$$\rm_{0}$=71 km s$^{-1}$ Mpc$^{-1}$ is assumed.

\newpage
\section{Abell\,449 and J0349$+$7511}
\label{sec:J0349}

The radio source J0349$+$7511 (central position R.A.: $\rm 03^{h}49^{m}16\fs28$ 
decl.: $75\degr11\arcmin22\farcs0$, J2000.0) is located in the cluster of galaxies Abell\,449 
(central position R.A.: $\rm 03^{h}49^{m}33^{s}$ decl.: $75\degr11\arcmin11\arcsec$, J2000.0). 
The cluster's redshift (determined as an average from four galaxies, but without specifying 
their identification) is $z=0.0803$~\citep{Postman1985}, which yields a distance of about 360\,Mpc. 
The angular size of Abell\,449 is about $23\farcm5$~\citep{Struble1982}, which corresponds to $\sim2.1$\,Mpc. 
According to~\cite{Flin2006}, the cluster has an unimodal structure with a central dominant region. 
Its richness and distance classes in the Abell Catalog are 1~and~4, respectively~\citep{Postman1992}. 
The cluster was classified together with Abell\,527 as one supercluster. The redshift 
of the parent galaxy of J0349$+$7511 is not known. However, as it is located near the center of 
Abell\,449 and its optical magnitude (R$\sim$ 14.3 mag; taken from the USNO-B catalog, \citealt{Monet2003}) 
is only slightly higher than three other neighboring galaxies, we presume that the galaxy actually belongs to the cluster. 
The angular size of J0349$+$7511, as measured along both diffuse lobes, is about $14\farcm5$, which corresponds 
to the linear size of $\sim$1.3\,Mpc. This size, much larger than that of radio galaxies mentioned in the introduction, 
makes it a quite unusual case among dying radio galaxies found to date.

\section{Radio Data}
\label{sec:data}

In order to perform a radio analysis of J0349$+$7511, we used data from 38\,MHz to 5\,GHz.
Most of the maps and measurements are taken from archives but the crucial 613\,MHz images come 
from dedicated observations with the GMRT. The common flux density scale of our measurements is 
that of ~\cite{Baars1977}.

\subsection{GMRT Observations}

The observations of the source at 613\,MHz were made with GMRT in the standard manner in 2014 November.
At the beginning, the primary calibrator, 3C48, was observed. Then each observation of the target-source was 
alternated with observations of the phase calibrator, 3C468.1 or 3C147. The collected data were automatically 
flagged with the use of the FLAGCAL software pipeline \citep{Chengalur2012, Chengalur2013} and then reduced 
and calibrated using the NRAO Astronomical Image Processing System (AIPS) software package\footnote{http://www.aips.nrao.edu}. Self-calibration was performed several times to improve the quality of the images. 
The resolution and the rms noise of the map are $7\farcs0\times5\farcs1$ and 0.03 mJy beam$^{-1}$, respectively. Subsequently, the original (u, v) data set was tapered at 15 kilo-lambda to make the diffuse structure more visible. 
The resulting $11\arcsec\times11\arcsec$ resolution map is presented in Figure~\ref{fig1}. Its rms is 
0.05\,mJy\,beam$^{-1}$. In addition, to imagine well the compact structures, not contaminated by extended emission, 
we obtained a third map which excluded data below 3 kilo-lambda. 
Three images of compact radio emission sources are shown in Figure~\ref{fig2} (bottom panel). In order 
to compare the radio emission at 613\,MHz with the low-resolution ($\sim1\arcmin$) WENSS map, we 
tapered the original (u, v) data set again, but in this case at 2.9 kilo-lambda, obtaining a map with rms 
noise of 0.19\,mJy\,beam$^{-1}$ and angular resolution of about $45\arcsec$. The $3\sigma$ contour of this 
low-resolution map is shown with a thick line in Figure~\ref{fig1}.

\subsection{Archival Observations}

Radio data pertaining to J0349$+$7511 are available in the modern radio sky surveys, e.g., the 
Very Large Array (VLA) Low-frequency Sky Survey redux (VLSSr; 74\,MHz;~{\citealt{Lane2014}), 
the WENSS (326\,MHz) and the NVSS (1400\,MHz). The two separated objects of this source 
are available in the Eight Cambridge Survey (8C; 38\,MHz;~\citealt{Hales1995}). We have also 
taken the VLA archival data at 1477\,MHz from $\sim8$ minute snapshot observations performed 
in May 1988 (project ID AO\,0082), when the antennas were in DnC configuration and reduced them 
in a standard way using the AIPS package. J0349$+$7511 is not detected in the Green Bank 
4.85\,GHz sky survey \citep{Condon1989}. 

\section{Infrared and X-ray data}
\label{sec:multi}

The archival Wide-field Infrared Survey Explorer satellite (WISE; \citealt{Wright2010}) 
observations at 3.4\,{$\mu$}m (W1), 4.6\,{$\mu$}m (W2), 12\,{$\mu$}m (W3), 22\,{$\mu$}m 
(W4) in the vicinity of the cluster are available. The data related to the dominant galaxy, 
J034915.47$+$751121.9, found in the WISE catalog, give the following magnitudes: 
W1=$12.438\pm0.024$, W2=$12.375\pm0.024$, W3=$11.613\pm0.202$ and W4=$9.139$. 
Using the $\rm W1-W2$ and $\rm W2-W3$ colors, it is possible to assess both the galaxy's 
and AGN's type (e.g., \citealt{Gurkan2014, Wright2010}). The values calculated for 
J0349$+$7511 indicate that the parent galaxy is of elliptical type and the central source is 
a low-excitation radio galaxy.

In the X-ray images showing the vicinity of the Abell\,449 cluster of galaxies in the Roentgen Satellite 
(ROSAT) catalog of faint sources \citep{Voges2000} there appear three structures visible in the energy 
range 0.5-2.0\,keV, 1RXS\,J034915.2$+$750647, 1RXS\,J034948.7$+$751027, 1RXS\,J034953.7$+$751634, 
which have a count rate of $2.13\pm0.83$, $2.33\pm0.96$ and $1.86\pm0.73$\,cts\,s$^{-1}$, respectively. 
The central positions of these sources in respect to the radio structures are marked with white crosses 
in Figure~\ref{fig1}. Future X-ray observations, however, should clarify if any of these 
objects are located within the cluster. Anyway, the absence of strong X-ray emission would 
suggest that Abell\,449 is rather a low-mass system. Although cluster radio halos and relics usually reside within X-ray luminous massive galaxy aggregates (e.g., \citealt{Giovannini2009}), there are some exceptions to the rule (see e.g., \citealt{Giovannini2011}).

\section{Results}
\label{sec:results}

\subsection{Source Morphology and Spectral Index}

Figures~\ref{fig1}--~\ref{fig3} show the vicinity of J0349$+$7511 at different radio frequencies. 
The distinct radio sources within this cluster have been labelled: L1 and L2 mark the north-western 
and the south-eastern lobe, respectively; B1 and B2 mark the extended background sources; C1 is the 
radio core of J0349$+$7511 and C2 is probably a background source with no optical counterpart. 
The flux-densities for different regions along with their spectral indices are presented in 
Table~\ref{tab:fluxes}. 

The radio images show a double-lobed radio source that is axially symmetric. It resembles the giant 
radio galaxy J0200$+$4049 with possible relic lobes described by \cite{Godambe2009}. The two roundish 
diffuse lobes (L1 and L2) with no distinct hotspots do not seem to be connected with any prominent 
background/foreground galaxy. The L2 lobe is a brighter one and the flux-density ratio of the lobes 
at 613\,MHz is 1.3. The extended sources B1 and B2 are probably background FRII-type (\citealt{Fanaroff1974}) 
radio galaxies. There is a north-south extended central structure between the two lobes, which in 
the high-resolution images shows two resolved components C1 and C2. The northern component, where 
the dominant galaxy of this cluster is situated, has an optical counterpart. There are some hints that 
this component is radio-variable at 1400\,MHz (see Table~\ref{tab:fluxes}). Therefore, C1 seems to be the 
radio core of J0349$+$7511. Most of the flux density of the lobes (L1 and L2) is emitted at low 
frequencies, while their flux at high frequencies (e.g., NVSS) is faint, if detectable at all 
(see Figure~\ref{fig2}). Accordingly, these structures are apparently aged with spectral index 
$\sim1.5\pm0.1$ between 74 and 326\,MHz. The spectra of the central sources (C1 and C2) and the objects 
B1, B2 are flatter than those of the lobes.

\begin{figure*}[htb]
\begin{center}
\epsscale{1}
\includegraphics[scale=0.9]{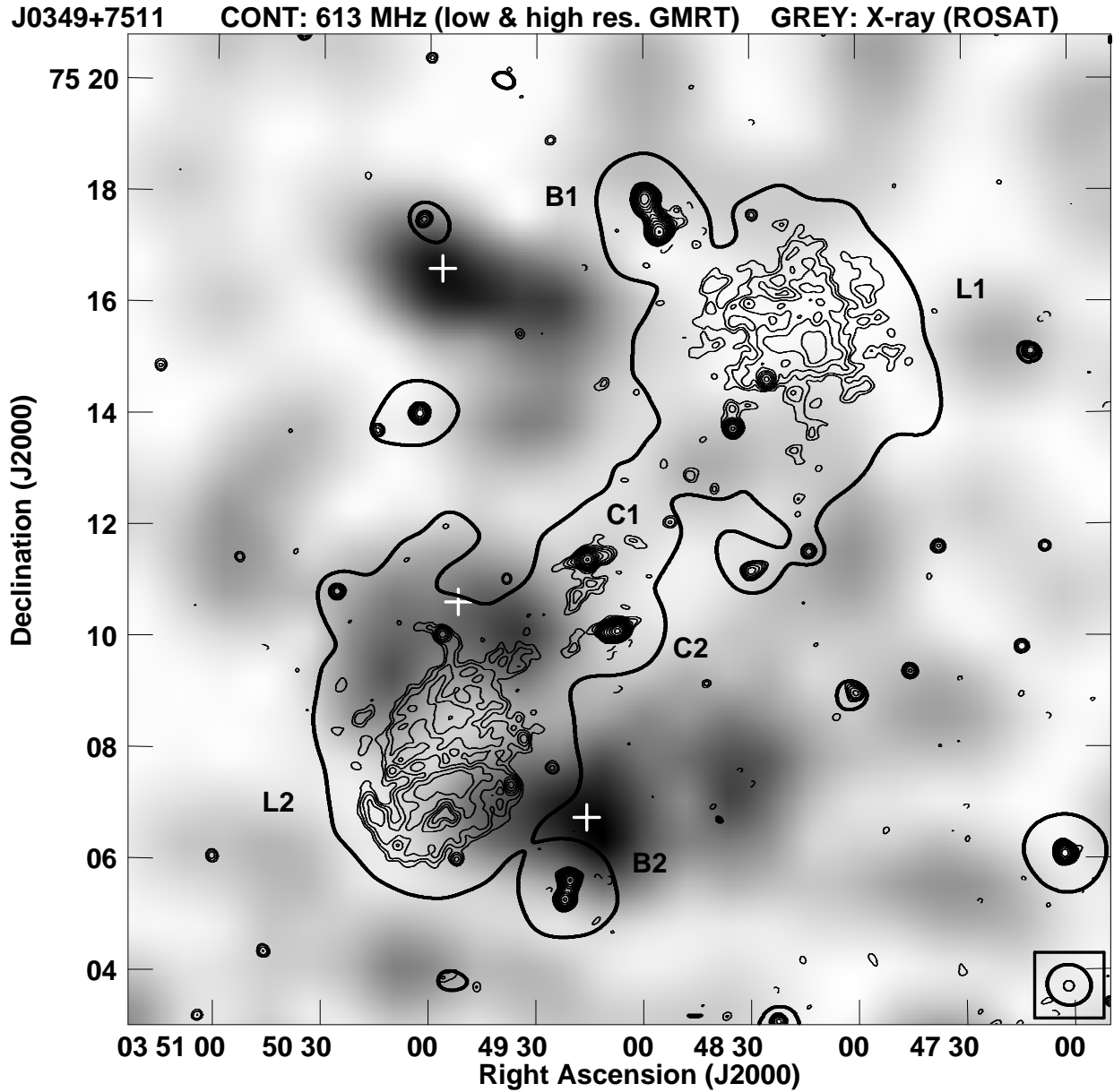}
\caption
{Ultra-steep spectrum radio galaxy J0349$+$7511. The indistinct outlines in gray-scale are the 
0.5-2.0\,keV ROSAT image with overlaid 613\,MHz GMRT thin contours (u, v)-tapered at 15\,kilo-lambda, 
starting from 0.2\,mJy\,beam$^{-1}$ and spaced by factors of $\sqrt{2}$ as well as the 613\,MHz GMRT 
thick contour (u, v)-tapered at 2.9\,kilo-lambda at level 0.65\,mJy\,beam$^{-1}$. 
The labels L1, L2, B1, B2, C1, C2 for different regions are given (for details, see the text). 
The resolving level of the high- and low-resolution GMRT radio images is given by an ellipse in 
the bottom right corner. The white crosses mark the position of the X-ray ROSAT sources.
}
\label{fig1}
\end{center}
\end{figure*}

Figure~\ref{fig3} presents the spectral index map across J0349$+$7511 and the close objects between 326\,MHz 
and 613\,MHz. The map was created from the GMRT and WENSS images with AIPS task COMB. The GMRT map had been 
earlier (u, v)-tapered at 2.9\,kilo-lambda, resized, and convolved with tasks HGEOM and CONVL, so that its 
geometry and resolution were similar to the WENSS image. There is no evident asymmetry of the spectrum 
of the lobes, which is steep with a mean value of $1.81\pm0.62$ and $2.12\pm0.58$ for the L1 and L2 lobe, 
respectively. The apparent steepening can be seen toward the northern edge of the eastern lobe. We note that 
the very steep radio spectra seen at some points at the outskirts of the structure ($\alpha < -3$) may be due 
to some missing flux in the 613\,MHz interferometric map. The central part of the source has a much flatter 
spectrum than the lobes. The C2 component has a spectral index of $0.87\pm0.31$, while the mean spectral index 
of the presumable source core (C1) is $0.93\pm0.40$. There are regions on both sides of the core, possibly 
remains of some backflow, which have a much steeper spectrum of about $2.25\pm0.43$ and $1.63\pm0.24$ 
on the western and eastern side, respectively. The radio core reveal probably a recent jet activity cycle. 
After the jet activity in a fading radio source stops, it can appear again. In this case, the emission 
breaks out once more and the spectrum of the new central structure becomes much flatter than those of the 
diffuse regions. However, the spectrum of a core of a young radio source can be steep ($\alpha\gtrsim0.8$) 
at high frequencies ($\gtrsim1$\,GHz). Such sources are called compact steep-spectrum or gigahertz 
peaked-spectrum radio galaxies. Indeed, a small fraction of radio galaxies with relics/dying lobes feature 
active radio cores. Those objects are called recurrent activity radio galaxies 
(for details, see \citealt{Saikia2009}). In the case of J0349$+$7511, we are dealing just with such a source.

\subsection{Source energetics}

J0349$+$7511 is a low-luminosity radio source with the 613\,MHz power of
$\rm P_{613 MHz} = 2.57\times10^{24}$\,W Hz$^{-1}$. The source volume has been calculated assuming 
a spherical shape for the radio lobes, and a rough estimation gives $\rm V=4.8\times10^{72}$\,cm$^{3}$. 
For deriving the synchrotron age of the particles within the relic lobes, the strength of the magnetic 
field is required. In estimating it, we rely on the equipartition arguments, and in our calculations
we followed the formalism of \cite{BeckKrause2005}. With the help of the \cite{BeckKrause2005} package named
BFIELD, assuming the ratio of energy between protons and electrons to be unity and the spectral index
$\alpha=1.87$, we obtained the magnetic field strength of the lobes as $B_{\rm eq}=$0.113$\pm$0.015\,nT, 
comparable to the value obtained for the remnant radio galaxy J18282048$+$4914428 \citep{Brienza2016}. The absence of 
any apparent jets and hotspots suggests that particle acceleration within the lobes no longer takes place. 
In order to estimate the total synchrotron age $\tau_{\rm rad}$ of a relic radio galaxy, 
one can follow the model (usually labelled as KGJP; \citealt{Komissarov1994}), assuming two phases of its life, 
i.e. a period of continuous particle injection, followed by ageing. This method was 
used by e.g., \cite{Murgia2011}, \cite{Shulevski2015}, \cite{Brienza2016} 
who estimated ages of relic radio galaxies. After their birth, radio galaxies are supposed 
to be fuelled at a constant rate (continuous injection (CI) phase, \citealt{Kardashev1962}) 
for a duration of $\tau_{\rm CI}$. In this phase, the source's radio spectrum changes 
at a break frequency $\nu_{\rm br}$. At the time $\tau_{\rm CI}$, the power  
supply from the nucleus is switched off and the relic phase of 
duration $\tau_{\rm RE}$ begins. A new break at higher frequency $\nu_{\rm brh}$
then appears. This second break is related to the first one by
$\frac{\nu_{\rm brh}}{\nu_{\rm br}}=(\frac{\tau_{\rm rad}}{\tau_{\rm RE}})^{2}=(1+\frac{\tau_{\rm CI}}{\tau_{\rm RE}})^{2}$. The KGJP model is described by four parameters. Unfortunately, having flux density measurements at four frequencies only (and one upper limit from the NVSS), 
the model fitting would be affected by significant errors. Therefore, we  decided to fit 
the spectrum of J0349$+$7511 assuming only one phase of its life, with the model of \cite{Jaffe1973}. 
Using the SYNAGE package \citep{Murgia1996}, we performed a fit to the lobes' spectrum 
and computed the break frequency $\nu_{\rm br}=$ 0.45$\pm$0.12\,GHz, above which the radio spectrum steepens
from the injected power-law slope. The $\nu_{\rm br}$ is related to the spectral age through
\begin{equation}
\tau_{\rm rad}=50.3\frac{B_{\rm eq}^{1/2}}{B_{\rm eq}^{2}+B^{2}_{\rm CMB}}\{\nu_{\rm br}(1+z)\}^{-1/2}\, {\rm Myr},
\label{eqn_specage}
\end{equation}
 
\noindent
where $B_{\rm CMB}=0.37$\,nT is the magnetic field strength equivalent to the cosmic microwave background 
radiation at the redshift of the source. The obtained $B_{\rm CMB}$ value is $\sim$ 3 times larger than 
the source's minimum energy magnetic field. Therefore, the energy losses in this relic radio galaxy are
probably dominated by the inverse Compton scattering. The estimated mean synchrotron age of the lobes' 
particles is about $160\pm20$\,Myr. The estimated age, however, is rather just a lower limit of the total 
source age, and for a more accurate estimation it is necessary to have flux-density measurements at additional frequencies. According to \cite{BeckKrause2005}, the estimated energy density
within the lobes is $\rm u_{eq}=7.7\pm2.2\times10^{-21}$\,J cm$^{-3}$. The energy
density value is about two times lower than those obtained for J18282048$+$4914428 \citep{Brienza2016}. 
$\rm u_{eq}$ is directly related to the lobe pressure, $p_{\rm eq} = (\gamma-1)\times u_{\rm eq}$, were 
$\gamma$ is the ratio of specific heats. In the case of ultra-relativistic gas, $\gamma=\frac{4}{3}$, 
and hence the pressure (estimated by assuming energy equipartition and pressure balance conditions) is 
$p_{\rm eq}=25.7 \pm 7.3 \times10^{-15}$\,dyn cm$^{-2}$. The corresponding particle density of the ICM
can be estimated as $n_{\rm eq}=u_{\rm eq}/3kT$. Here $k$ is the Boltzmann constant and $T$ is the temperature of 
the ICM (in K). For temperatures of $10^{7}$\,K, it gives $n_{\rm eq}= 18.6 \pm 5.3 \times10^{-6}$\,cm$^{-3}$. 

\begin{figure*}[htb]
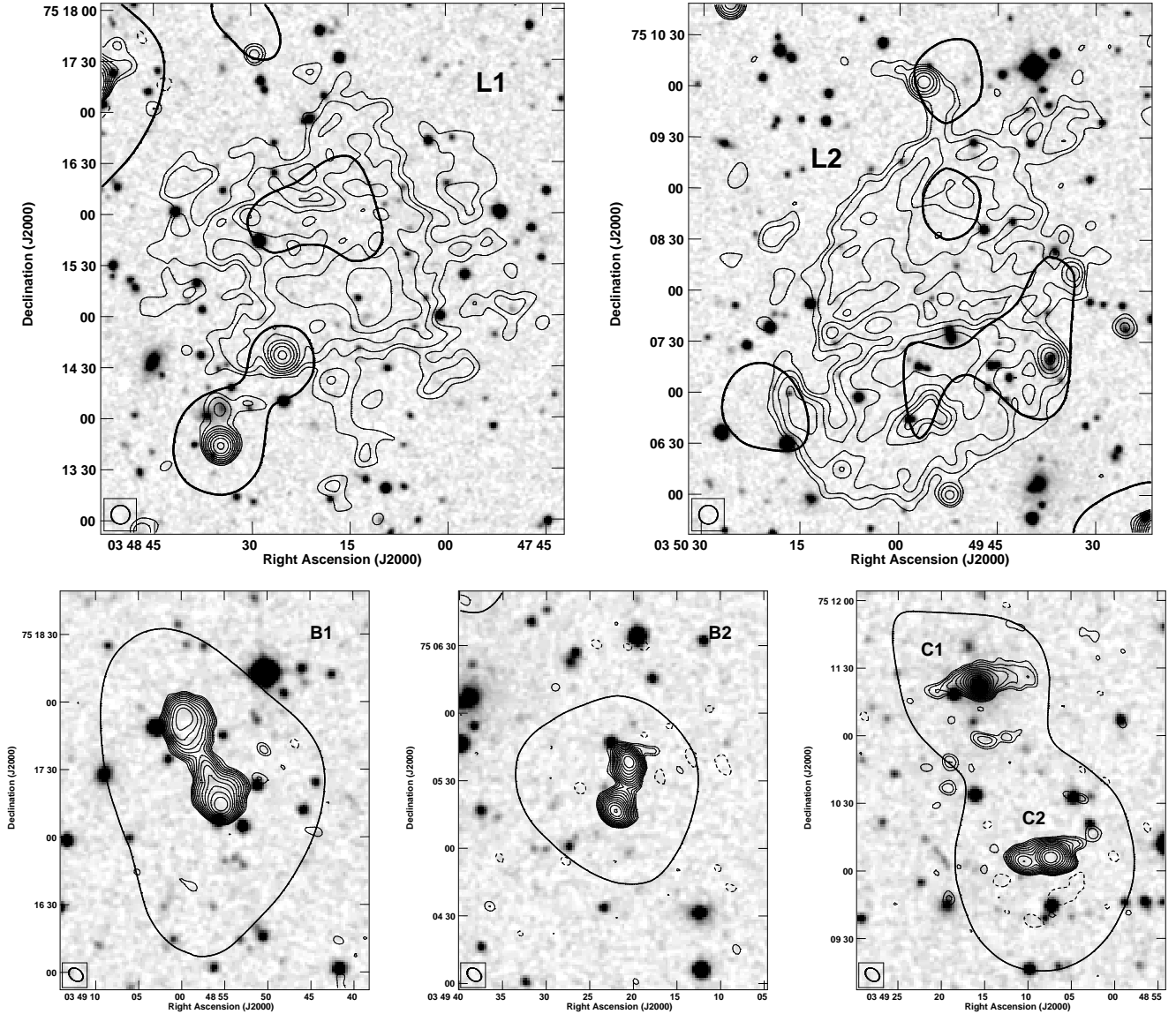

\center
\includegraphics[scale=0.45]{fig2a.eps}
\;
\includegraphics[scale=0.45]{fig2b.eps}
\\ \;
\includegraphics[scale=0.3]{fig2c.eps}
\;
\includegraphics[scale=0.3]{fig2d.eps}
\;
\includegraphics[scale=0.3]{fig2e.eps}

\caption
{Enlargement of the individual parts of J0349$+$7511 and the close background radio sources. 
Top panels: the vicinity of the L1 lobe (left panel) and L2 lobe (right panel) superimposed 
on gray-scale optical R-band DSS image. The 613\,MHz GMRT contours (thin lines), starting from 
0.2\,mJy\,beam$^{-1}$ and spaced by factors of $\sqrt{2}$, and the 1400\,MHz NVSS contour (thick line) 
at level of 1.2\,mJy\,beam$^{-1}$ are given. The GMRT map was (u,v)-tapered at 15\,kilo-lambda. 
The size of the 11\arcsec$\times$11\arcsec\, GMRT beam is indicated by an ellipse in the bottom 
left corner of the images. Bottom panels: images of the two extended background sources B1 (left panel) 
and B2 (middle panel), as well as of the central sources (right panel) C1 (the radio core of J0349$+$7511) 
and C2 (a background source), superimposed on gray-scale optical R-band DSS images. The 613\,MHz GMRT contours 
(thin lines) starting from 0.12\,mJy\,beam$^{-1}$, spaced by factors of $\sqrt{2}$ and the 1400\,MHz NVSS contour 
(thick line) at level 1.2\,mJy\,beam$^{-1}$, are given. The size of the 6\farcs2$\times$4\farcs6 GMRT 
beam is indicated by the ellipse in the bottom left corner of the images.
}
\label{fig2}
\end{figure*}

\begin{table*}[htb]
\begin{center}
\caption
{\small Flux densities and spectral indices for the different components of J0349$+$7511 and the close 
background sources.}
\begin{tabular}{c c c c c c c c}
\hline
\hline
               & \multicolumn{6}{c}{Survey/Telescope}							    & Spectr.      \\
\cline{2-7}
Region         & 8C             & VLSSr       & WENSS         & GMRT         & NVSS         & VLA           & index        \\
               & 38 (MHz)       & 74 (MHz)    & 326 (MHz)     & 613 (MHz)    & 1400 (MHz)   & 1477 (MHz)    & $\pm$error   \\
\hline              
               & \multicolumn{6}{c}{Flux density $\pm$ error (mJy)}                                         &              \\
\cline{2-7}
L1             &  13500$\pm1400$& 2270$\pm$390& 330.0$\pm20.0$ & 67.3$\pm3.5$ & \dots        & \dots        & 1.86$\pm$0.17\\
L2             &  18300$\pm1800$& 4270$\pm$510& 463.0$\pm26.0$ & 87.5$\pm4.5$ & \dots        & \dots        & 1.88$\pm$0.17\\
B1             &  \dots         & \dots       &  55.8$\pm6.1 $ & 41.4$\pm2.1$ & 23.9$\pm1.8$ & 19.9$\pm1.4$ &0.70$\pm0.09$\\
B2             &  \dots         & \dots       &  38.5$\pm5.1 $ & 22.2$\pm1.1$ &  8.3$\pm1.3$ &  5.4$\pm0.7$ &1.29$\pm0.20$\\
C1             &  \dots         & \dots       &  \dots         &  8.8$\pm0.5$ &  4.4$\pm0.4$ &  2.7$\pm0.4$ &0.99$\pm0.27$\\
C2             &  \dots         & \dots       &  \dots         & 21.4$\pm1.1$ & 12.0$\pm0.6$ & 11.4$\pm0.4$ &0.71$\pm0.01$\\
C1+C2          &  \dots         & \dots       &  72.2$\pm6.9$  & 30.2$\pm1.5$ & 16.4$\pm0.7$ & 14.1$\pm0.6$ &0.95$\pm0.12$\\
\hline
\hline
\end{tabular}
\tablecomments
{The original flux density of the  8C survey (RBC scale; \citealt{Roger1973}) and the VLSSr survey were multiplied
by the factors of 0.83 and 0.9 respectively, to be consistent with the \cite{Baars1977} scale. Other
fluxes were measured in the obtained images or taken from the NVSS catalog. The sources C1 and C2 could not be
resolved in the WENSS image. The flux density errors are estimated as follows: the 8C survey data are assumed
to have the errors as taken from the literature; for the other data, we decided to adopt 5\% calibration errors;
in the case of an extended source, a noise term related to the size of its structure (the product of map noise and square
root of the number of beams per structure) is added, and the overall error is a root mean square of the calibration and
noise errors. The spectral index values are obtained by linear fitting to the presented flux densities.}	
\label{tab:fluxes}
\end{center}
\end{table*}

\begin{figure}[htb]
\begin{center}
\includegraphics[scale=0.45]{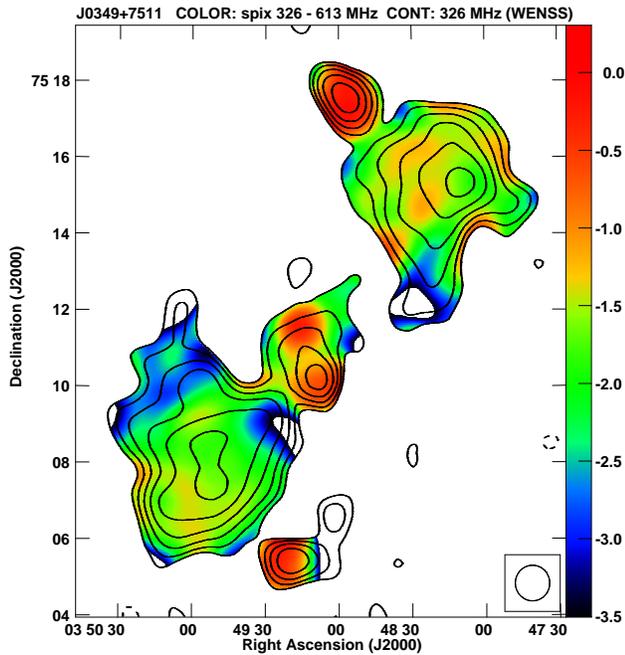}
\caption
{
Spectral index image of J0349$+$7511. Color scale represents spectral index map between 326\,MHz (WENSS) 
and 613\,MHz (GMRT) with overlaid 326\,MHz WENSS contours, starting from level of 
7\,mJy\,beam$^{-1}$ and spaced by factors of $\sqrt{2}$. The size of the beam is indicated by the 
ellipse in the bottom right corner of the image.
}
\label{fig3}
\end{center}
\end{figure}

\section{Conclusion}
\label{sec:conclusion}

We identified a new ultra-steep 1.3\,Mpc giant radio galaxy in Abell\,449. The source J0349$+$7511
is well visible at low radio frequencies, while hardly detectable above 1.4 GHz. Therefore, J0349$+$7511
is an excellent target for further study with the new radio telescopes operating in the meter wavelengths, 
i.e. the Low Frequency Array and the Long Wavelength Array. The estimated mean synchrotron age of the extended 
diffuse relic lobes is about 160\,Myr. The parent galaxy located symmetrically between the lobes seems to be 
starting a new jet activity cycle. 

\section{Acknowledgments}
We thank the anonymous Referee, whose comments and suggestions allowed us to significantly improve 
this article. We thank Prof. Jayaram Chengalur for providing us with the FLAGCAL software and Dr. Matteo Murgia for access to the SYNAGE software. We thank the staff of GMRT, who made these observations possible. GMRT is run by the National Centre for Radio Astrophysics of the Tata Institute of Fundamental Research. M.J. and D.H. acknowledge support by the Polish National Science Centre grant No. 2013/09/B/ST9/00599 and a scholarship of Marian Smoluchowski Research Consortium Matter Energy Future from KNOW funding, respectively.

\end{document}